\documentstyle[12pt,epsf]{article}

\renewcommand{\bar}[1]{\overline{#1}}

\begin{document}
\begin{flushright}
SLAC--PUB--7272\\
August 1996
\end{flushright}
\bigskip\bigskip

\thispagestyle{empty}
\flushbottom

\bibliographystyle{unsrt}
\def\Journal#1#2#3#4{{#1} {\bf #2}, #3 (#4)}
\def\NCA{\em Nuovo Cimento}
\def\NIM{\em Nucl. Instrum. Methods}
\def\NIMA{{\em Nucl. Instrum. Methods} A}
\def\NPB{{\em Nucl. Phys.} B}
\def\PLB{{\em Phys. Lett.}  B}
\def\PRL{\em Phys. Rev. Lett.}
\def\PRD{{\em Phys. Rev.} D}
\def\ZPC{{\em Z. Phys.} C}
\def\st{\scriptstyle}
\def\sst{\scriptscriptstyle}
\def\mco{\multicolumn}
\def\epp{\epsilon^{\prime}}
\def\vep{\varepsilon}
\def\ra{\rightarrow}
\def\ppg{\pi^+\pi^-\gamma}
\def\vp{{\bf p}}
\def\ko{K^0}
\def\kb{\bar{K^0}}
\def\al{\alpha}
\def\ab{\bar{\alpha}}
\def\be{\begin{equation}}
\def\ee{\end{equation}}
\def\bea{\begin{eqnarray}}
\def\eea{\end{eqnarray}}
\def\CPbar{\hbox{{\rm CP}\hskip-1.80em{/}}}

\centerline{\large\bf NONPERTURBATIVE RENORMALIZATION}
\centerline{{\large\bf OF QED IN LIGHT-CONE QUANTIZATION}\footnote{
\baselineskip=14pt
     Work supported in part by the Minnesota Supercomputer
Institute through grants of computing time and by the 
Department of Energy, contract 
     DE--AC03--76SF00515.}}
\vspace{22pt}

\centerline{\bf J. R. Hiller}
\vspace{8pt}
\centerline{\it Department of Physics, University of Minnesota}
\centerline{\it Duluth, Minnesota~~55812}
\vspace{10pt}
\centerline{and}
\vspace{10pt}
  \centerline{\bf Stanley J. Brodsky}
\vspace{8pt}
  \centerline{\it Stanford Linear Accelerator Center}
  \centerline{\it Stanford University, Stanford, California 94309}
\vspace*{0.9cm}

\begin{center}
ABSTRACT
\end{center}
   
As a precursor to work on QCD, we study the dressed electron in QED
nonperturbatively.  The calculational scheme uses an invariant mass
cutoff, discretized light-cone quantization, a Tamm--Dancoff
truncation of the Fock space, and a small photon mass.
Nonperturbative renormalization of the coupling and electron mass
is developed.
\vfill
\centerline{Paper submitted to the}
\centerline{1996 Annual Divisional Meeting (DPF96) of the}
\centerline{Division of Particle and Fields of the}
\centerline{American Physical Society}
\centerline{University of Minnesota, Minneapolis, Minnesota}
\centerline{10--15 August 1996}
\vfill
\newpage
\section{Introduction}

We are in the process of studying dressed fermion states in a gauge
theory.  To give the work specific focus, we concentrate on the
nonperturbative calculation of the anomalous moment of the
electron.~\cite{Hiller} This is not intended to be competitive with
perturbative calculations.~\cite{Kinoshita} Instead it is an
exploration of nonperturbative methods that might be applied to QCD and
that might provide a response to the challenge by
Feynman~\cite{Feynman} to find a better understanding of the anomalous
moment.

The methods used are based on light-cone quantization~\cite{Reviews}
and on a number of approximations.  Light-cone coordinates provide for
a well-defined Fock state expansion.  We then approximate the expansion
with a Tamm--Dancoff~\cite{TammDancoff} truncation to no more than two
photons and one electron. The Fock-state expansion can be written
schematically as $\Psi=\psi_0|e\rangle+\psi_1|e\gamma\rangle
+\psi_2|e\gamma\gamma\rangle$. The eigenvalue problem for the wave
functions $\psi_i$ and the bound-state mass $M$ becomes a coupled set
of three integral equations.  To construct these equations we use the
Hamiltonian $H_{\rm LC}$ of Tang {\em et al}.~\cite{Tang} The anomalous
moment is then calculated from the spin-flip matrix element of the plus
component of the current.~\cite{BrodskyDrell} The regulator is an
invariant-mass cutoff $\sum_i (P^+/p_i^+)\left(m_i^2+p_{\perp
i}^2\right)\leq\Lambda^2$. Additional approximations and assumptions
are a nonzero photon mass of $m_e/10$, a large coupling of
$\alpha=1/10$, and use of numerical methods based on discretized
light-cone quantization (DLCQ).~\cite{Reviews}

\section{Renormalization}

We renormalize the electron mass and couplings differently in each Fock
sector, as a consequence of the Tamm--Dancoff
truncation.~\cite{SectorDependent} The bare electron mass in the
one-photon sector is computed from the one-loop correction allowed by
the two-photon states. We then require that the bare mass in the
no-photon sector be such that $M^2=m_e^2$ is an eigenvalue.

The three-point bare coupling $e_0$ is related to the physical coupling
$e_R$ by
 $e_0(\underline{k}_i,\underline{k}_f)
    =Z_1(\underline{k}_f)e_R/
          \sqrt{Z_{2i}(\underline{k}_i)Z_{2f}(\underline{k}_f)}$,
where $\underline{k}_i=(k_i^+,{\bf k}_{\perp i})$ is the initial
electron momentum and $\underline{k}_f$ the final momentum.  The
renormalization functions $Z_1(\underline{k})$ and
$Z_2(\underline{k})=|\psi_0|^2$ are generalizations of the usual
constants. The amplitude $\psi_0$ must be computed in a basis where
only allowed particles appear.

The function $Z_1$ can be fixed by considering the proper part of the
transition amplitude $T_{fi}$ for photon absorption by an electron at
zero photon momentum
($\underline{q}=\underline{k}_f-\underline{k}_i\rightarrow 0$):
$T_{fi}^{\rm proper}=V_{fi}/Z_1(\underline{k}_f)$, where $V_{fi}$ is
the elementary three-point vertex. The transition amplitude can be
computed from $T_{fi}=\psi_0\langle\Psi|V|i\rangle$, in which
$|\Psi\rangle$ is the dressed electron state and
$\psi_0=\sqrt{Z_{2f}(\underline{k}_f)}$. The proper amplitude is then
obtained from $T_{fi}^{\rm proper}=T_{fi}/(Z_{2i}Z_{2f})$, where the
$Z_2$'s remove the disconnected dressing of the electron lines.

Thus the solution of the eigenvalue problem for only one state can be
used to compute $Z_1$.  Full diagonalization of $H_{\rm LC}$ is not
needed.  Because $Z_1$ is needed in the construction of $H_{\rm LC}$,
the eigenvalue problem and the renormalization conditions must be
solved simultaneously.

Most four-point graphs that arise in the bound-state problem are log
divergent.  To any order the divergences cancel if all graphs are
included, but the Tamm--Dancoff truncation spoils this. For a
nonperturbative calculation we need a counterterm
$\sim\lambda(p_i^+,p_f^+)\log\Lambda$ that includes infinite chains of
interconnected loops.  The function $\lambda$ might be fit to Compton
amplitudes.~\cite{MustakiPinsky} Thus we need to be able to handle
scattering processes.

\section{Preliminary Results and Future Work}

\begin{figure}[htb]
\begin{center}
\leavevmode 
\epsfbox{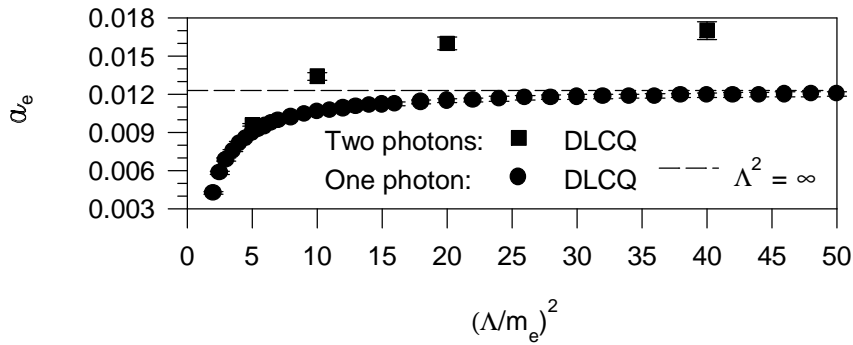}
\end{center}
{\caption[*]{Electron anomalous moment as a function of the cutoff
$\Lambda^2$, extrapolated from DLCQ calculations.  The photon mass is
$m_e/10$, and the coupling is 1/10.} 
\label{fig:a_e}} 
\end{figure}

Some preliminary results are given in Fig.~\ref{fig:a_e}.  In the
two-photon case there remain divergences associated with four-point
graphs.

The next step to be taken in this calculation is renormalization of the
four-point couplings, followed by numerical verification that all logs
have been removed.  Construction of finite counterterms that restore
symmetries will then be considered.  We can also consider photon zero
modes, Z graphs, and pair states.

\vspace{5pt}

\section*{Acknowledgments}
This work has benefited from discussions with R. J. Perry, St. D.
G{\l}azek, and \break G.~McCartor.

\end{document}